\documentclass[11pt,paper]{article}
\usepackage{jheppub}

\usepackage{float,slashed}
\usepackage[usenames,dvipsnames,table]{xcolor}
\usepackage{graphicx,amsmath,amssymb,multirow,array,bm,mathrsfs}
\usepackage{epsf,amsfonts,slashed}
\usepackage[numbers,sort&compress]{natbib}
\usepackage[export]{adjustbox}
\usepackage{tikz}
\usetikzlibrary{decorations.pathmorphing}
\usetikzlibrary{decorations.markings}
\usepackage{scalerel,stackengine} 
\usepackage{soul}
\usepackage[utf8]{inputenc}
\usepackage{hyperref}
\usepackage{tikz-feynman}

\title{Gravitationally dominated instantons and instability of dS, AdS and Minkowski spaces
}

\date{}

\author[a,b]{Viatcheslav F. Mukhanov,}
\author[c]{Yaron Oz,}
\author[d]{Alexander S. Sorin}

\affiliation[a]{Ludwig Maximilian University, Theresienstr. 37,
80333 Munich, Germany
 }
\affiliation[b]{Korea Institute for Advanced Study Seoul,
02455, Korea}
\emailAdd{mukhanov@physik.lmu.de}
\affiliation[c]{
School of Physics and Astronomy, Tel-Aviv University, Tel-Aviv 69978, Israel
 }
 \emailAdd{yaronoz@tauex.tau.ac.il}
\affiliation[d]{Center for Quantum Science and Technology, Tel-Aviv University, Tel-Aviv 69978, Israel
 } 
 \emailAdd{asorin@tauex.tau.ac.il}

\abstract
{We study the decay of the false vacuum in the regime where the quantum field 
theory analysis is not valid, since gravitational effects  become important.
This happens when 
the height of the barrier separating the false and the true vacuum
is large, and it has implications for the instability of de Sitter, Minkowski and anti-de
Sitter vacua. We carry out the calculations for a scalar field with a potential coupled to gravity, 
and work within the thin-wall approximation, where the bubble wall is thin compared to the size of the bubble. 
We show that the false de Sitter vacuum is unstable, independently of
the height of the potential and the relative depth of the true vacuum
compared to the false vacuum. The false Minkowski and anti-de Sitter
vacua can be stable despite the existence of a lower energy true vacuum.
However, when the relative depth of the true and false vacua exceeds a critical value,
which depends on the potential of the false vacuum and the height
of the barrier, then the false Minkowski and anti-de Sitter vacua become unstable.  We calculate the
probability for the decay of the false de Sitter, Minkowski and anti-de
Sitter vacua,  as a function of the parameters characterizing the 
field potential.}

\begin{document}

\maketitle



\section{Introduction}

The decay of the false vacuum in quantum field theory (QFT), i.e.
the transition from a metastable state to a more stable state through
quantum tunneling, is a non-perturbative process that has significant
implications for cosmology and particle physics. The false vacuum
is a local minimum of the potential energy of a quantum field, but
it is not the lowest energy state, which is the true vacuum. The process
of the decay of the false vacuum involves the formation of bubbles
in which small regions of the true vacuum spontaneously form within
the false vacuum due to quantum fluctuations. Once a bubble of the
true vacuum has formed, it can expand if the energy density inside
the bubble is less than the energy density of the surrounding false
vacuum and the bubble wall separates the two vacua. 

The bubble grows
when the energy gained from the false vacuum outweighs the energy
cost of the bubble wall. As bubbles of true vacuum grow and merge,
they percolate the entire space, leading to a transition from the
false vacuum to the true vacuum. The decay of the false vacuum can
be analyzed using the semiclassical framework \cite{Kobzarev:1974cp,PhysRevD.15.2929}. 
In this framework, the decay
rate per unit volume and unit time is $\Gamma\sim Ae^{-S}$, where
$A$ is a dimensional coefficient that is inversely proportional to the
fourth power of the bubble radius,
and $S$ is the Euclidean action of the
bounce solution.

The aim of this work is to analyse the decay of the false vacuum when
gravity becomes important, and the QFT analysis is not valid. 
Our
analysis is performed in the framework of relativistic field theory
coupled to gravity and extends the analysis of \cite{Coleman:1980aw} to the regime
where gravity dominates the sub-barrier transition.
This
happens when the height of the barrier separating the false and the
true vacua is large, and it has important implications for the instability
of de Sitter (dS), Minkowski and anti-de Sitter (AdS) vacua.

We will work within
the thin-wall approximation \cite{Kobzarev:1974cp,PhysRevD.15.2929}, 
where the bubble wall is thin compared
to the size of the bubble, and show that: (i) the false dS
vacuum is always unstable, regardless of the height of the potential
and the relative depth of the true vacuum compared to the false vacuum,
(ii) the false Minkowski vacuum and the AdS vacuum can
be stable despite the existence of a true vacuum with lower energy.
However, if the relative depth of the vacua exceeds a critical value,
which depends on the value of the potential in the false vacuum and the height
of the barrier, then the false vacuum in these two cases becomes unstable
and decays into a true vacuum. We calculate the probability for the
decay of the false dS, Minkowski and AdS vacua as
a function of the parameters characterizing the field potential. 

The paper is organized as follows. In section II, we will consider
the structure and range of validity of gravitationally dominated instantons
in the thin-wall approximation. In section III we will analyse the
instabilities of dS, Minkowski and AdS vacua in the regime, where
gravity plays an important role. Section IV will be devoted to a discussion.

\section{Gravitationally dominated instantons}

\begin{figure}[h!]
    \centering
    \includegraphics[width=0.5\textwidth]{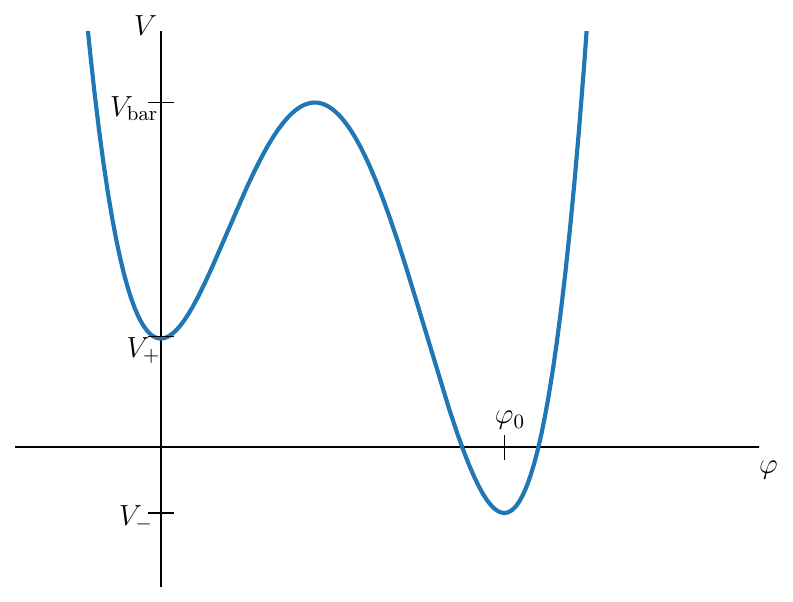}
    \caption{A scalar field potential $V(\varphi)$ that has a local minimum false vacuum
at $\varphi=0$, where $V\left(\varphi=0\right)=V_+$, 
which is separated by a barrier of height $V_{bar}$ from the true vacuum at $\varphi= \varphi_{0}>0$,
with $V\left(\varphi_{0}\right)=V_-$.}
\label{potential}
\end{figure}

We will consider a scalar field with a potential shown in Fig.\ref{potential},
which is coupled to gravity.

\subsection{The Newtonian approximation}

To gain insight into the importance of gravity for the decay of the
false vacuum, we first perform the field theory analysis in Minkowski
space, taking into account the leading gravitational corrections
in the Newtonian approximation. Let us assume that the scalar field
potential $V(\varphi)$ has a local minimum corresponding to false
vacuum at $\varphi=0$, where $V\left(\varphi=0\right)=V_+=\varepsilon$,
which is separated from the true Minkowski vacuum at $\varphi=\varphi_{0}>0$,
with $V\left(\varphi_{0}\right)=V_-=0$, by a barrier of height $V_{bar}$.
The false vacuum is unstable and will decay by the formation of critical
bubbles, whose size $\varrho_{b}$ can be estimated using the energy
conservation. Working without gravity, and equating the total energy
released during the transition, $4\pi\varepsilon\varrho_{b}^{3}/3$,
with the surface energy of the bubble wall, $4\pi\sigma\varrho_{b}^{2}$,
where $\sigma$ is the surface tension of the bubble, we obtain \cite{Kobzarev:1974cp,PhysRevD.15.2929}:
\begin{equation}
4\pi\varepsilon\varrho_{b}^{3}/3=4\pi\sigma\varrho_{b}^{2}\implies\varrho_{b}=\frac{3\sigma}{\varepsilon}\ .\label{eq:1a}
\end{equation}

The gravitational corrections to (\ref{eq:1a}) are parametrized by
an effective dimensionless gravitational coupling $g_{eff}=\kappa M/\varrho$,
where $\kappa\equiv8\pi G$ with $G$ being the gravitational Newton
constant, $\varrho$ and $M$ are the characteristic length and mass
in the problem. In our case, $\varrho$ is the bubble size $\varrho_{b}$,
and $M$ is the total energy released during the transition and the
surface energy of the bubble wall. The first order gravitational corrections
to the energy balance equation (\ref{eq:1a}) come from the negative
gravitational energy contributions and in the Newtonian approximation
the balance equation reads: 
\begin{equation}
\frac{4\pi}{3}\varepsilon\varrho_{b}^{3}-\frac{3}{5}\frac{\kappa(4\pi\varepsilon\varrho_{b}^{3}/3)^{2}}{8\pi\varrho_{b}}=4\pi\sigma\varrho_{b}^{2}-\frac{\kappa\left(4\pi\sigma\varrho_{b}^{2}\right)^{2}}{16\pi\varrho_{b}}\ ,\label{eq:2a}
\end{equation}
where we have neglected higher order terms in $g_{eff}$. Equation
(\ref{eq:2a}) shows the potential importance of gravity for bubble
formation, but its range of validity is limited to $g_{eff}\ll1$.
Looking at the LHS of (\ref{eq:01a}), this means that equation (\ref{eq:2a})
is valid provided: 
\begin{equation}
\frac{\kappa\frac{4\pi}{3}\varepsilon\varrho_{b}^{3}}{\varrho_{b}}\ll1\implies\varrho_{b}\ll\frac{1}{\sqrt{\kappa\varepsilon}}\ .\label{eq:01a}
\end{equation}
This is the regime in which $\varrho_{b}\ll R_{+}$, where $R_{+}=\sqrt{3/(\kappa\varepsilon)}$
is the dS radius corresponding to the false vacuum.
If we take into account the condition $g_{eff}\ll1$ in the RHS of
(\ref{eq:01a}) we get 
\begin{equation}
\varrho_{b}\ll\frac{1}{\kappa\sigma}\ ,
\end{equation}
which can be related in the thin-wall approximation to the parameters
of the potential since $\sigma\sim\varphi_{0}V_{bar}^{1/2}$ (see,
e.g. \cite{Mukhanov:2005sc}). As we will see, the interesting phenomena
caused by gravity occur when $g_{eff}\sim O(1)$, i.e. $\varrho_{b}\sim R_{+}$
or $\varrho_{b}\sim1/(\kappa\varphi_{0}V_{bar}^{1/2})$, where the
analysis requires general relativity, and this will be carried out in the
next section. 
Note, that the dimensionless quantum gravity expansion parameter is $\kappa^2\epsilon$, which will be taken to be much smaller than one in our analysis.
\subsection{Instantons basics}

As argued in \cite{PhysRevD.15.2929}, the most important contribution
to the sub-barrier transition between the false and the true vacua
is due to the $O(4)$ Euclidean instantons, for which the value of
the scalar field $\varphi$ depends on the Euclidean time $\tau$
and the spatial radial coordinate $r$ in the combination $\tau^{2}+r^{2}$.
When the gravitational effects can be neglected, the background metric
for these instantons is the flat Euclidean metric. The gravitational
backreaction of the instantons on the metric that preserves the $O(4)$
symmetry, results in a conformally flat metric\footnote{We use the signature (-,+++) to perform the analytical continuation
of the Minkowski metric to Euclidean space by Wick rotation of Minkowski
time $t_{M}=-i\tau$.}: 
\begin{equation}
ds^{2}=f^{2}\left(\sqrt{\tau^{2}+r^{2}}\right)\left(d\tau^{2}+dr^{2}+r^{2}d\Omega^{2}\right),\label{eq:5aa}
\end{equation}
where $d\Omega^{2}=d\theta^{2}+\sin^{2}\theta d\varphi^{2}$. 

Define: 
\[
\tau=\bar{\xi}\sin\chi,\quad r=\bar{\xi}\cos\chi\ ,
\]
where $\infty>\bar{\xi}\geq0$ and $\pi/2\geq\chi\geq-\pi/2$, and
then the metric reads: 
\begin{equation}
ds^{2}=f^{2}\left(\bar{\xi}\right)\left[d\bar{\xi}^{2}+\bar{\xi}^{2}\left(d\chi^{2}+\cos^{2}\chi d\Omega^{2}\right)\right]\ .\label{eq:7aa}
\end{equation}
Introducing:
\begin{equation}
d\xi=f\left(\bar{\xi}\right)d\bar{\xi},\qquad\varrho=f\left(\bar{\xi}\right)\bar{\xi}\ ,\label{eq:8aa}
\end{equation}
we recast (\ref{eq:7aa}) as\footnote{For the convenience of the reader, we will use the same notations
almost everywhere as in the paper on gravitational effects for
instantons by Coleman and De Luccia \cite{Coleman:1980aw}.}: 
\begin{equation}
ds^{2}=d\xi^{2}+\varrho^{2}\left(\xi\right)\left(d\chi^{2}+\cos^{2}\chi d\Omega^{2}\right)\ .\label{eq:9aa}
\end{equation}
The off-shell Euclidean action for a scalar field $\varphi\left(\xi\right)$
with potential $V\left(\varphi\right)$ propagating in the background
metric (\ref{eq:9aa}) reads: 
\begin{equation}
S_{E}=2\pi^{2}\int d\xi\left[\varrho^{3}\left(\frac{1}{2}\dot{\varphi}^{2}+V\right)+\frac{3}{\kappa}\left(\varrho^{2}\ddot{\varrho}+\varrho\left(\dot{\varrho}^{2}-1\right)-(\varrho^{2}\dot{\varrho})^{\centerdot}\right)\right]\ ,\label{eq:10aa}
\end{equation}
where the dot denotes a derivative with respect to $\xi$, and the
last term is a total derivative, corresponding to York-Gibbons-Hawking
boundary term \cite{PhysRevLett.28.1082,Gibbons:1976ue}.

The variation of the action
with respect to $\varphi$ yields its field equation: 
\begin{equation}
\ddot{\varphi}+3\frac{\dot{\varrho}}{\varrho}\dot{\varphi}-V_{,\varphi}=0\ ,\label{eq:11aa}
\end{equation}
where $V_{,\varphi}=dV/d\varphi$, while the variation with respect
to $\varrho$ gives: 
\begin{equation}
2\varrho\ddot{\varrho}+\dot{\varrho}^{2}-1=-\kappa\varrho^{2}\left(\frac{1}{2}\dot{\varphi}^{2}+V\right)\ .\label{eq:12aa}
\end{equation}
Multiplying (\ref{eq:12aa}) by $\dot{\varrho}$ and using (\ref{eq:11aa})
to express $\varrho^{2}\dot{\varrho}\dot{\varphi}^{2}$ in the RHS
in terms of $\varrho$ and the derivative of $\left(\dot{\varphi}^{2}/2-V\right)$
we get: 
\begin{equation}
\left(\varrho\left(\dot{\varrho}^{2}-1\right)\right)^{\centerdot}=\frac{\kappa}{3}\left(\varrho^{3}\left(\frac{1}{2}\dot{\varphi}^{2}-V\right)\right)^{\centerdot}\ .\label{eq:12aaa}
\end{equation}
This can be integrated to obtain: 
\begin{equation}
\dot{\varrho}^{2}=1+\frac{\kappa}{3}\varrho^{2}\left(\frac{1}{2}\dot{\varphi}^{2}-V\right)\ ,\label{eq:14aa}
\end{equation}
where we set the integration constant to zero in accordance with the
$0-0$ component of the Einstein equations. Inserting (\ref{eq:14aa})
in (\ref{eq:12aa}), we get: 
\begin{equation}
\ddot{\varrho}=-\frac{1}{3}\kappa\varrho\left(\dot{\varphi}^{2}+V\right)\ .\label{eq:16aa}
\end{equation}

To find the instanton action that dominates the tunneling rate, we
need to solve the equations (\ref{eq:11aa}) and (\ref{eq:14aa})
with the appropriate boundary conditions. Using these equations, we
can rewrite the on-shell action in a form that is more convenient
for calculating the instanton action. Substituting (\ref{eq:14aa})
in (\ref{eq:10aa}), the on-shell action simplifies to: 
\begin{equation}
S_{E}=2\pi^{2}\int d\xi\left[\varrho^{3}\dot{\varphi}^{2}+\frac{3}{\kappa}\left(\varrho^{2}\ddot{\varrho}-(\varrho^{2}\dot{\varrho})^{\centerdot}\right)\right]\ ,\label{eq:17aa}
\end{equation}
which using (\ref{eq:16aa}) can be written as: 
\begin{equation}
S_{E}=-2\pi^{2}\int d\xi\left(\varrho^{3}V+\frac{3}{\kappa}(\varrho^{2}\dot{\varrho})^{\centerdot}\right)\ .\label{eq:18aa}
\end{equation}
The decay rate of the false vacuum per unit volume per unit time can
be estimated as \cite{Coleman:1980aw}: 
\begin{equation}
\Gamma\sim\varrho_{b}^{-4}e^{-(S_{E}^{f}-S_{E}^{i})},\label{col1}
\end{equation}
where $\varrho_{b}$ is the size of the resulting critical bubble.
$S_{E}^{i}$ and $S_{E}^{f}$ are the corresponding Euclidean actions
for the initial false vacuum and for the final configuration, where
the true vacuum has been created inside the critical bubble.

Note, that so far all the equations were exact, and we have not made
any approximation. In order to obtain the quantitative results
for general potentials $V$, we will have to make approximations.
In the next section we will consider the thin-wall approximation in
the presence of gravity. We will also denote the Euclidean action $S_{E}$ without the subscript $E$.

\subsection{The thin wall approximation}

Let us consider a false vacuum with potential $V_{+}$, located at
$\varphi=0$, and separated from the true vacuum of depth $V_{-}<V_{+}$
at $\varphi_{0}$ by the barrier with positive height $V_{bar}\gg\left|V_{+}\right|$.
Depending on $V_{-}$, the false dS vacuum decays either into
a true dS vacuum with a lower potential or into Minkowski or AdS vacuum.
The decay occurs via instantons, which lead to the formation of critical
bubbles of size $\varrho_{b}$. We assume that the thickness of the
bubble wall $\delta\varrho_{b}$ is much smaller than $\varrho_{b}$.
This corresponds to the case, where the second term (``friction'')
in the equation (\ref{eq:11aa}) can be neglected in the leading order
approximation. 

To determine the range of validity of this approximation,
we consider the first integral of the equation (\ref{eq:11aa}) 
\begin{equation}
\frac{1}{2}\dot{\varphi}^{2}-V=-\int\frac{3\dot{\varrho}}{\varrho}\dot{\varphi}^{2}d\xi\ .\label{eq:19aa}
\end{equation}
In the thin-wall approximation, the integral on the RHS of (\ref{eq:19aa}),
to which only the bubble wall contributes where $\dot{\varphi}^{2}$
does not vanish, must be small compared to each individual term on
the LHS of (\ref{eq:19aa}) within the wall. Let us first assume that
this condition is fulfilled, calculate the radius of the critical
bubble, and determine for which potentials the thin-wall approximation
can be applied. If we neglect the integral on the RHS in (\ref{eq:19aa}),
then we obtain: 
\begin{equation}
E\equiv\frac{1}{2}\dot{\varphi}^{2}-V\approx-V_{+}\ .\label{eq:19abc}
\end{equation}

Assuming that $\left|V_{+}\right|\ll V_{bar}$, we can substitute
$V\approx\dot{\varphi}^{2}/2$ within the wall into equation (\ref{eq:16aa})
and integrate over the wall to obtain: 
\begin{equation}
\dot{\varrho}_{+}-\dot{\varrho}_{-}=-\frac{1}{2}\kappa\sigma\varrho_{b}\ ,\label{eq:26aa}
\end{equation}
where $\dot{\varrho}_{-}$ and $\dot{\varrho}_{+}$ refer to the true
and false vacuum immediately before and after the thin wall, and $\sigma$
is the surface tension defined as:
\begin{equation}
\sigma\equiv\int\dot{\varphi}^{2}d\xi\ .\label{eq:21aa}
\end{equation}
For $\left|V_{+}\right|\ll V_{bar}$, $\sigma$ can be estimated as: 
\begin{equation}
\sigma=\int\dot{\varphi}d\varphi\simeq\int\sqrt{2V}d\varphi\sim\sqrt{V_{bar}}\varphi_{0}\ .\label{eq:25aa}
\end{equation}
Using (\ref{eq:14aa}) we have: 
\begin{equation}
\dot{\varrho}_{+}=\pm\sqrt{1-\frac{\kappa V_{+}}{3}\varrho_{b}^{2},}\qquad\dot{\varrho}_{-}=+\sqrt{1-\frac{\kappa V_{-}}{3}\varrho_{b}^{2}}\ .\label{eq:27aa}
\end{equation}
Note, that the space inside the resulting bubble expands and $\dot{\varrho}_{-}$
is therefore always positive, while outside the bubble $\dot{\varrho}_{+}$
can be either positive or negative, depending on the height of the
barrier $V_{bar}$ and the other parameters characterising the field
potential.

Solving (\ref{eq:26aa}) gives an expression for the size of the critical
bubble 
\begin{equation}
\varrho_{b}=\left[\frac{R_{+}^{2}}{1+\left(\frac{\varepsilon}{3\sigma}\left(1-\alpha\right)R_{+}\right)^{2}}\right]^{1/2},\label{eq:28aa}
\end{equation}
where $R_{+}^{2}=3/\kappa V_{+}$, $\varepsilon\equiv V_{+}-V_{-}>0$
is difference between the potential heights in the false and true
vacuum, respectively, and $\alpha$ is a dimensionless parameter:
\begin{equation}
\alpha\equiv\frac{3\kappa\sigma^{2}}{4\varepsilon}\ .\label{eq:31aaa-1}
\end{equation}
Inserting (\ref{eq:28aa}) into (\ref{eq:27aa}) we get: 
\begin{equation}
\dot{\varrho}_{+}=\pm\frac{\varepsilon}{3\sigma}\left|1-\alpha\right|\varrho_{b},\quad\dot{\varrho}_{-}=\frac{\varepsilon}{3\sigma}\left(1+\alpha\right)\varrho_{b}\ .\label{eq:31aa}
\end{equation}

Note, that the above calculations apply to both positive and negative
$V_{+}$ and $V_{-}$, and can be used for the instability analysis
of both dS and AdS spaces. For the false dS vacuum, $V_{+}>0$ and
$R_{+}$ is the dS radius. When $\alpha>1$, we have to put a negative
sign in front of the square root in $\dot{\varrho}_{b^{+}}$ in (\ref{eq:27aa})
to fulfill (\ref{eq:26aa}) and therefore: 
\begin{equation}
\dot{\varrho}_{+}=\,\text{sgn}(1-\alpha)\sqrt{1-\frac{\kappa V_{+}}{3}\varrho_{b}^{2},}\qquad\dot{\varrho}_{-}=+\sqrt{1-\frac{\kappa V_{-}}{3}\varrho_{b}^{2}}\ .\label{eq:33a}
\end{equation}
Next, let us consider to which potentials the thin-wall approximation
can be applied. As we can see from (\ref{eq:16aa}), $\dot{\varrho}$
always decreases as we go through the bubble wall, and therefore the
integral on the RHS in (\ref{eq:19aa}) is bounded by 
\begin{equation}
\int\frac{3\dot{\varrho}}{\varrho}\dot{\varphi}^{2}d\xi<\int\frac{3\dot{\varrho}_{-}}{\varrho_{b}}\dot{\varphi}^{2}d\xi=\varepsilon+\frac{3\kappa\sigma^{2}}{4}\ , \label{eq:33aa}
\end{equation}
and it must be much smaller than $V_{bar}$. Using $\sigma\sim\sqrt{V_{bar}}\varphi_{0}$,
we find that the conditions for applicability of the thin-wall approximation
are met when: 
\begin{equation}
\frac{\varepsilon}{V_{bar}}\ll1,\qquad\kappa\varphi_{0}^{2}\ll1\ .\label{eq:34c}
\end{equation}

For the instantons with $\alpha>1$, the expanding dS solution inside
the critical bubble is always matched with a contracting dS branch,
describing the false vacuum outside the bubble. In this case, the
value of $\dot{\varrho}$ should vanish inside the bubble wall, and
its sign must change from positive to a negative inside the wall.
As can be seen from (\ref{eq:14aa}), this would be impossible if
the energy $E$ (\ref{eq:19abc}) remains exactly the same. However,
due to the friction term in (\ref{eq:11aa}), the energy $E$ changes
slightly. As can be seen from (\ref{eq:14aa}), this slight change
is sufficient to violate the energy conservation law (\ref{eq:19abc})
by the amount $\Delta E\sim3/(\kappa\varrho_{b}^{2})$ required to change
the sign of the expansion rate $\dot{\varrho}$ within the wall, which
is much smaller than $V_{bar}$ for the instantons with $\alpha>1$,
if $\kappa\varphi_{0}^{2}\ll1.$ Further, equation (\ref{eq:19abc})
which was used in the derivation (\ref{eq:26aa}) for the thin-wall
bubble, where $\kappa\varphi_{0}^{2}\ll1$ applies, is therefore also
fulfilled in the leading order.

\section{Instability of dS, AdS and Minkowski spaces}

In this section we will analyse the instabilities of dS, Minkowski
and AdS spaces, as well as the decay of the false vacuum in the regime
where gravity is important and the QFT analysis is not valid.

\subsection{The decay probability of the dS vacuum}

To determine the decay rate of the dS false vacuum, we need to calculate
the change in Euclidean on-shell action during the sub-barrier transition.
In a false vacuum, $\dot{\varphi}=0$, and the equation (\ref{eq:14aa}) simplifies
to: 
\begin{equation}
\dot{\varrho}^{2}=1-\frac{\varrho^{2}}{R_{+}^{2}} \ ,\label{eq:34aa}
\end{equation}
where $R_{+}=\sqrt{3/(\kappa V_{+})}$, and is solved by: 
\begin{equation}
\varrho\left(\xi\right)=R_{+}\sin\left(\frac{\xi}{R_{+}}\right)\ .\label{eq:35aa}
\end{equation}
When $\xi$ changes from $0$ to $\pi R_{+}$, the radius of the Euclidean
dS instanton first increases, reaches its maximal value $\varrho=R_{+}$
at $\xi=\frac{1}{2}\pi R_{+}$, and then shrinks to zero.

Let us first consider the case in which $\alpha<1$. According to
(\ref{eq:33a}), in the final configuration where the critical bubble
is created, we must match the expanding solution describing a true
vacuum with a potential $V_{-}$ inside the bubble with an expanding
dS solution corresponding to the false vacuum outside the bubble.
The radius of the critical bubble $\varrho_{b}$ is smaller than $R_{+}$,
and the wall is located at $\xi_{b}<\frac{1}{2}\pi R_{+}.$ The contribution
of the false vacuum in the final configuration to the total action
can be calculated by subtracting from the initial false vacuum action
$S^{i}$ the action for the small bubble with radius $\varrho_{b}$,
with a cosmological constant corresponding to the false vacuum: 
\begin{equation}
S_{+}^{f}=S^{i}+2\pi^{2}\intop_{0}^{\varrho_{b}}\frac{V_{+}\varrho^{3}}{\sqrt{1-\kappa V_{+}\varrho^{2}/3}}d\varrho+\frac{6\pi^{2}}{\kappa}\varrho_{b}^{2}\dot{\varrho}_{+}=S^{i}+\frac{12\pi^{2}}{\kappa^{2}V_{+}}\left(1-\dot{\varrho}_{+}^{3}\right)\ .\label{eq:37abc}
\end{equation}
In the derivation, we used (\ref{eq:18aa}) for the action, and replaced the integration
over $\xi$ by the integration over $\varrho$ using (\ref{eq:14aa})
with the positive sign for the square root.

The action for the true vacuum expanding solution inside the bubble
is calculated in the same way: 
\begin{equation}
S_{-}^{f}=-2\pi^{2}\intop_{0}^{\varrho_{b}}\frac{V_{-}\varrho^{3}}{\sqrt{1-\kappa V_{-}\varrho^{2}/3}}d\varrho-\frac{6\pi^{2}}{\kappa}\varrho_{b}^{2}\dot{\varrho}_{-}=-\frac{12\pi^{2}}{\kappa^{2}V_{-}}\left(1-\dot{\varrho}_{-}^{3}\right)\ .\label{eq:38aa}
\end{equation}
Finally, the contribution of the thin wall is obtained by integrating
(\ref{eq:17aa}) over the wall, and taking into account the definition
of the tension in (\ref{eq:21aa}): 
\begin{equation}
S_{w}^{f}=2\pi^{2}\sigma\varrho_{b}^{3}\ .\label{eq:39aa}
\end{equation}
Combining (\ref{eq:37abc}), (\ref{eq:38aa}) and (\ref{eq:39aa}),
we obtain the total change in action between the initial and final
configurations, which includes the critical bubble of the true vacuum
separated from the false vacuum by the thin wall: 
\begin{align}
 & S^{f}-S^{i}=2\pi^{2}\sigma\varrho_{b}^{3}+\frac{12\pi^{2}}{\kappa^{2}}\left[\frac{1-\dot{\varrho}_{+}^{3}}{V_{+}}-\frac{1-\dot{\varrho}_{-}^{3}}{V_{-}}\right]\nonumber \\
 & =\frac{12\pi^{2}}{\kappa^{2}}\left[\frac{1-\dot{\varrho}_{+}}{V_{+}}-\frac{1-\dot{\varrho}_{-}}{V_{-}}\right]=\frac{2\pi^{2}\sigma\varrho_{b}^{3}}{\left(1+\dot{\varrho}_{+}\right)\left(1+\dot{\varrho}_{-}\right)}\ ,\label{eq:40ab}
\end{align}
where we used (\ref{eq:26aa}) to simplify the final result.

Performing a similar calculation for the case $\alpha>1$, and taking
into account that in this case we have to match the expanding true
vacuum inside the bubble with the contracting false dS vacuum and
accordingly take a negative sign for $\dot{\varrho}_{+}$, we get
exactly the same expression. Thus, for each $\alpha>0$, the decay
rate of the false vacuum is determined by (\ref{col1}), where the
size of the critical bubble $\varrho_{b}$ is given in (\ref{eq:28aa}),
and the change in the action (\ref{eq:40ab}) is calculated with $\dot{\varrho}_{+}$
and $\dot{\varrho}_{-}$ from (\ref{eq:33a}).

\begin{figure}[h!]
    \centering
    \includegraphics[width=0.5\textwidth]{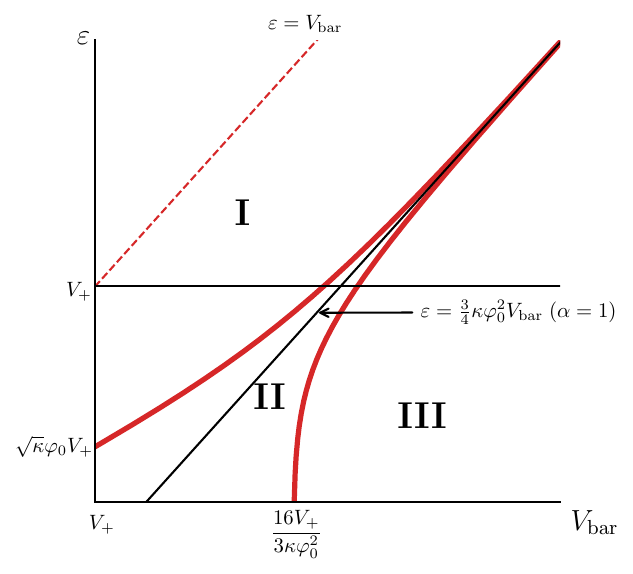}
    \caption{A plot of the plane $\varepsilon-V_{bar}$ on a logarithmic scale 
    with $\varepsilon>0$ and $V_{bar}>V_{+}$. Region I refers to the field theory instantons.
Regions II and III describe gravitationally dominated instantons,  with
a radius of the order of the dS radius $R_{+}$ (II), and 
with a radius much smaller than $R_{+}$ (III). }
\label{regimes}
\end{figure}

As follows from (\ref{eq:28aa}), the size of the critical bubble
never exceeds the radius of the false dS vacuum, and $\varrho_{b}=R_{+}$
is reached for $\alpha=1$. Given $V_{+}$ and $\varphi_{0}$, we
aim to find out how the decay rate depends on the height of the potential
barrier $V_{bar}$ separating false and true vacuum, and on the difference
between the depths of the vacua $\varepsilon=V_{+}-V_{-}>0$. It is
convenient to illustrate the results using the plane $\varepsilon-V_{bar}$
in Fig.\ref{regimes}, where region I refers to the field theory
instantons, and regions II and III describe gravitationally dominated
instantons, with a radius of the order of the dS radius $R_{+}$ (II),
and with radius much smaller than $R_{+}$ (III), respectively.

\subsubsection{Field theory instantons}

\noindent {}~ {}~ {}~ {}~ {\bf  \textit{Region} I} in Fig.\ref{regimes}: If we assume that
$\alpha\ll1$ and 
\begin{equation}
\frac{\varepsilon}{3\sigma}(1-\alpha)R_{+}\gg1\ ,\label{eq:50a}
\end{equation}
or, equivalently, 
\begin{equation}
\varepsilon\gg\frac{3}{4}\kappa\sigma^{2}\left(1+\frac{4}{\kappa\sigma R_{+}}\right)\ ,\label{eq:51-1}
\end{equation}
then the radius of the critical bubble (\ref{eq:28aa}) reads: 
\begin{equation}
\varrho_{b}\simeq\frac{3\sigma}{\varepsilon}\ .\label{eq:42abc}
\end{equation}
Using $\sigma\simeq\varphi_{0}V_{bar}^{1/2}$ and $\varepsilon\ll V_{bar}$
(\ref{eq:34c}), it follows that for a given height of the barrier
$V_{bar}$, the condition (\ref{eq:51-1}) can be rewritten as: 
\begin{equation}
V_{bar}\gg\varepsilon\gg\frac{3}{4}\kappa\varphi_{0}^{2}V_{bar}\left(1+\sqrt{\frac{V_{+}^{*}}{V_{bar}}}\right)\ ,\label{eq:43ad}
\end{equation}
where 
\begin{equation}
V_{+}^{*}\equiv\frac{16V_{+}}{3\kappa\varphi_{0}^{2}}\ .\label{eq:43adc}
\end{equation}
When this condition is met, we can neglect gravity and field theory
instantons provide a valid approximation to the decay of the false
vacuum. The false dS vacuum decays via these instantons to another
dS vacuum with a smaller potential or to Minkowski space if $\varepsilon\leq V_{+}$.
For $V_{bar}\gg\varepsilon\geq V_{+}$, the dS vacuum decays to an
AdS space, whose radius $R_{AdS}=\sqrt{3/(\kappa\left|V_{-}\right|)}$
is always larger than the radius of the bubble.

Finally, we can set $\dot{\varrho}_{b^{+}}=\dot{\varrho}_{b^{-}}\simeq1$
in (\ref{eq:40ab}), and obtain the known result for the field theory
instanton action 
\begin{equation}
S^{f}-S^{i}=\frac{\pi^{2}\sigma\varrho_{b}^{3}}{2}=\frac{27\pi^{2}}{2}\frac{\sigma^{4}}{\varepsilon^{3}}\ .\label{eq:45ac}
\end{equation}

\subsubsection{Gravitationally dominated instantons}

\noindent {}~ {}~ {}~ {}~ {\bf \textit{Region} II} in Fig.\ref{regimes}: If 
\begin{equation}
\frac{\varepsilon}{3\sigma}\left|1-\alpha\right|R_{+}\leq1\ ,\label{eq:46an}
\end{equation}
then, as follows from (\ref{eq:28aa}) the radius of the critical
bubble is of the order of the dS radius: 
\begin{equation}
\varrho_{b}\simeq R_{+}\ .\label{eq:46abc}
\end{equation}
In this case gravity plays a significant role. The value of $\alpha$
in (\ref{eq:46an}) can be either less than or greater than one. Therefore,
the inequality (\ref{eq:46an}) can be rewritten as: 
\begin{equation}
\frac{3}{4}\kappa\varphi_{0}^{2}V_{bar}\left(1+\sqrt{\frac{V_{+}^{*}}{V_{bar}}}\right)>\varepsilon>\frac{3}{4}\kappa\varphi_{0}^{2}V_{bar}\left(1-\sqrt{\frac{V_{+}^{*}}{V_{bar}}}\right)\ .\label{eq:47an}
\end{equation}
If $V_{bar}\ll V_{+}^{*}$, both $\left(1+\dot{\varrho}_{+}\right)$
and $\left(1+\dot{\varrho}_{-}\right)$ in (\ref{eq:40ab}) are of
order one, and the action that determines the transition rate from
the false dS vacuum to the true dS vacuum with $V_{+}>V_{-}>V_{+}\left(1-\sqrt{V_{bar}/V_{+}^{*}}\right)$
is given by: 
\begin{equation}
S^{f}-S^{i}\simeq O\left(1\right)\pi^{2}\sigma R_{+}^{3}\simeq O\left(1\right)\pi^{2}\left(\frac{V_{bar}}{V_{+}^{*}}\right)^{1/2}\frac{1}{\kappa^{2}V_{+}}\ .\label{eq:47ac}
\end{equation}

For $V_{bar}\gg V_{+}^{*}$, the dS vacuum can decay via instantons
of radius $R_{+}$ only into an AdS vacuum if $\varepsilon\simeq3\kappa\varphi_{0}^{2}V_{bar}/4$.
In this case 
\begin{equation}
1+\dot{\varrho}_{-}\simeq1+\sqrt{\frac{\varepsilon}{V_{+}}}\simeq\sqrt{\frac{3\kappa\varphi_{0}^{2}V_{bar}}{4V_{+}}}=\frac{1}{2}\kappa\sigma R_{+}\ \label{eq:47an-1}
\end{equation}
and as follows from (\ref{eq:40ab}) the action is: 
\begin{equation}
S^{f}-S^{i}\simeq\frac{4\pi^{2}}{\kappa}R_{+}^{2}=\frac{12\pi^{2}}{\kappa^{2}V_{+}}\ .\label{eq:55a}
\end{equation}

\noindent {}~ {}~ {}~ {}~ {\bf \textit{Region} III} in Fig.\ref{regimes}: Finally, let us consider
the case when $\alpha>1$, and 
\begin{equation}
\frac{\varepsilon}{3\sigma}(\alpha-1)R_{+}\gg1\ ,\label{eq:49an}
\end{equation}
i.e., 
\begin{equation}
\frac{3}{4}\kappa\varphi_{0}^{2}V_{bar}\left(1-\sqrt{\frac{V_{+}^{*}}{V_{bar}}}\right)\gg\varepsilon>0\ .\label{eq:49cn}
\end{equation}
This only applies if the barrier is sufficiently high, $V_{bar}>V_{+}^{*}$.
It then follows from (\ref{eq:28aa}) that 
\begin{equation}
\varrho_{b}\simeq\frac{4}{\kappa\sigma}\sim\frac{O(1)}{\kappa\varphi_{0}V_{bar}^{1/2}}\ .\label{eq:49ac}
\end{equation}
For such bubbles the negative contribution of gravity almost completely
compensates the energy released during the transition in the wall.
Taking into account that $\varrho_{b}\ll R_{+}$ for $V_{bar}\gg V_{+}^{*}$,
and that the sign of $\dot{\varrho}_{+}$ in (\ref{eq:33a}) is negative,
the action (\ref{eq:40ab}) reads: 
\begin{equation}
S^{f}-S^{i}=2\pi^{2}\sigma\varrho_{b}R_{+}^{2}\simeq\frac{8\pi^{2}}{\kappa}R_{+}^{2}=\frac{24\pi^{2}}{\kappa^{2}V_{+}}\ ,\label{eq:50ac}
\end{equation}
which is equal to the dS action of the false vacuum, but with an opposite
(positive) sign.

The decay rate of the false vacuum per unit time per unit volume is
given by: 
\begin{equation}
\Gamma\sim(\kappa\sigma)^{4}e^{-\frac{24\pi^{2}}{\kappa^{2}V_{+}}}\ ,\label{eq:51}
\end{equation}
and it does not depend on how deep the true vacuum is compared to
the false one. If $\varepsilon>V_{+}$, the dS vacuum decays to AdS
vacuum via bubbles, whose size is smaller than the radius of the AdS.
For $\varepsilon<V_{+}$ it decays to another dS true vacuum with
a smaller cosmological constant. The dependence on the height of the
barrier is in the pre-exponential factor. The higher the barrier is,
the smaller the critical bubbles are, and therefore the decay rate
increases as $V_{bar}^{2}$ with an increasing height of the barrier.
Thus, regardless of how high is the barrier between the false and
true vacuum, the false dS vacuum is always unstable and decays via
either the known field theory instantons or the gravitationally dominated
instantons, depending on the parameters of the potential.

\subsection{The instability of the Minkowski vacuum}

The Minkowski vacuum can decay if the true vacuum is an AdS vacuum
separated from Minkowski vacuum by the barrier of height $V_{bar}$.
The formulas describing this instability can be obtained by considering
the corresponding limiting case in the formulas derived above. In
particular, taking $R_{+}\rightarrow\infty$ in (\ref{eq:28aa}) leads
to the following expression for the radius of the critical bubble:
\begin{equation}
\varrho_{b}=\frac{3\sigma}{\varepsilon\left|1-\alpha\right|}\ .\label{eq:52}
\end{equation}
If $\alpha<1$, we obtain from (\ref{eq:40ab}) the following action
\begin{equation}
S^{f}-S^{i}=\frac{27\pi^{2}}{2}\frac{\sigma^{4}}{\varepsilon^{3}(1-\alpha)^{2}}\ .\label{eq:53}
\end{equation}
For $\alpha\ll1$, these results agree with the formulas describing
the field theory instantons (see (\ref{eq:42abc}) and (\ref{eq:45ac})).

The ratio of $\varrho_{b}$ to AdS radius $R_{AdS}\equiv\sqrt{3/(\kappa\left|V_{-}\right|)}$
is 
\begin{equation}
\frac{\varrho_{b}}{R_{AdS}}=\frac{2\sqrt{\alpha}}{1-\alpha}\ ,\label{eq:54}
\end{equation}
and therefore if 
\begin{equation}
\alpha=\frac{3\kappa\sigma^{2}}{4\varepsilon}\simeq\frac{3\kappa\varphi_{0}^{2}V_{bar}}{4\varepsilon}\ \label{eq:55}
\end{equation}
is of order one, the gravitational effects become important. In particular,
when $\alpha\rightarrow1$, the action tends to infinity. Note, that
the solution of (\ref{eq:14aa}) in Minkowski space $\varrho=+\xi$,
which corresponds to $\dot{\varrho}_{+}=1$ for the false vacuum,
covers the entire Euclidean space in the original coordinates (\ref{eq:7aa}),
where $\bar{\xi}$ changes in the range from zero to infinity. To
consider the case $\alpha>1$, in which the radius of the bubble is
given by (\ref{eq:49ac}) for $\alpha\gg1$, we should replace the
expanding coordinates in the false Minkowski vacuum by the contracting
coordinates. This in turn corresponds to an infinite change of the
action, and the resulting action is infinite. This can also be seen
from (\ref{eq:40ab}), where the denominator of $1+\dot{\varrho}_{+}$
vanishes. Consequently, the probability of the transitions vanishes
for $\alpha>1$.

Therefore, for a given height of the barrier $V_{bar}$, the Minkowski
vacuum is only unstable if the depth of the true AdS vacuum is large
enough, i.e. 
\begin{equation}
\left|V_{-}\right|=\varepsilon>\frac{3}{4}\kappa\varphi_{0}^{2}V_{bar}\ .\label{eq:56}
\end{equation}
Otherwise the Minkowski vacuum is stable, despite the existence of
the true AdS vacuum. The thin-wall approximation is of course only
valid for $\left|V_{-}\right|\ll V_{bar}$. The results obtained are
in complete agreement with the results for the dS false vacuum, if
we take the limit $V_{+}\rightarrow0$ in the formulas obtained in
the previous section.

\subsection{The transitions between the AdS vacua}

Let's look at the decay of the false AdS vacuum into another true
AdS vacuum. In this case $V_{+}$ and $V_{-}$ are negative, and we
assume that the magnitudes $\left|V_{+}\right|$ and $\left|V_{-}\right|$
are both much smaller than the positive height of the barrier $V_{bar}$.
Taking into account that $V_{+}<0$, we find from (\ref{eq:28aa})
the radius of the critical bubble: 
\begin{equation}
\varrho_{b}=\left[\frac{R_{+}^{2}}{\left(\frac{\varepsilon}{3\sigma}\left(1-\alpha\right)R_{+}\right)^{2}-1}\right]^{1/2}\ ,\label{eq:57}
\end{equation}
where now $R_{+}=\sqrt{3/(\kappa\left|V_{+}\right|)}$ is the radius
of the false AdS vacuum. It is always larger that the radius of the
true AdS vacuum $R_{-}=\sqrt{3/(\kappa\left|V_{-}\right|)}$, because
$\left|V_{-}\right|>\left|V_{+}\right|$.

It should be noted that the radius of the bubble is real when 
\begin{equation}
\frac{\varepsilon}{3\sigma}\left|1-\alpha\right|R_{+}>1\ .\label{eq:57a}
\end{equation}
Comparing this inequality with (\ref{eq:46an}), and taking into account
the fact that in this case we have to replace $V_{+}>0$ for dS with
$\left|V_{+}\right|$ for AdS, we find that the critical bubbles exist
only if either 
\begin{equation}
\varepsilon>\frac{3}{4}\kappa\varphi_{0}^{2}V_{bar}\left(1+\sqrt{\frac{\left|V_{+}^{*}\right|}{V_{bar}}}\right)\quad,\label{eq:57b}
\end{equation}
or
\begin{equation}
\varepsilon<\frac{3}{4}\kappa\varphi_{0}^{2}V_{bar}\left(1-\sqrt{\frac{\left|V_{+}^{*}\right|}{V_{bar}}}\right),\label{eq:57b1}
\end{equation}
where 
\begin{equation}
\left|V_{+}^{*}\right|\equiv\frac{16\left|V_{+}\right|}{3\kappa\varphi_{0}^{2}}\ .\label{eq:57c}
\end{equation}

The $\varepsilon-V_{bar}$ plane in the case of AdS is identical to
Fig.\ref{regimes}, if we replace $V_{+}$ by its magnitude of $\left|V_{+}\right|$,
when we consider the instability of the AdS false vacuum. As we have
shown above, in region II the equation for the critical bubble in
the case of AdS has no solutions, unlike the dS decay. Therefore,
we only need to calculate the probability of AdS decay in the regions
I and III, which correspond to the cases $\alpha<1$ and $\alpha>1$.

Equation (\ref{eq:14aa}) in the AdS case takes the form: 
\begin{equation}
\dot{\varrho}^{2}=1+\frac{\varrho^{2}}{R_{+}^{2}}\ ,\label{eq:58}
\end{equation}
and its expanding solution, 
\begin{equation}
\varrho\left(\xi\right)=R_{+}\sinh\left(\frac{\xi}{R_{+}}\right)\ ,\label{eq:59}
\end{equation}
does not cover the entire AdS false vacuum. Returning to the original
coordinates (\ref{eq:7aa}), we find that the $\bar{\xi}$ only changes
in the interval $0\leq\bar{\xi}\leq1$ for $0\leq\xi<\infty$. Therefore,
when considering transitions between AdS vacua, it is convenient
to work with the original coordinates (\ref{eq:7aa}). As can be easily
verified: 
\begin{equation}
\varrho\left(\bar{\xi}\right)=f\left(\bar{\xi}\right)\bar{\xi}=\frac{2R_{+}\bar{\xi}}{\left|1-\bar{\xi}^{2}\right|}\ ,\label{eq:60}
\end{equation}
where $0\leq\bar{\xi}\leq1$ describes the expanding branch of AdS,
and $1\leq\bar{\xi}\leq\infty$ its contracting branch, thus completely
covering the false AdS vacuum. Note, that $\bar{\xi}=1-0$ corresponds
to $\xi=+\infty$, while at $\bar{\xi}=1+0$ we have $\xi=-\infty$.
The metric (\ref{eq:7aa}) covering the entire AdS space thus becomes
\begin{equation}
ds^{2}=\frac{4R_{+}^2}{\left(1-\bar{\xi}^{2}\right)^{2}}\left[d\bar{\xi}^{2}+\bar{\xi}^{2}\left(d\chi^{2}+\cos^{2}\chi d\Omega^{2}\right)\right]\ ,\label{eq:61}
\end{equation}
where $0\leq\bar{\xi}<\infty$. Note, that $\bar{\xi}$ in this metric
can be shifted by an arbitrary constant $C$, $\bar{\xi}\rightarrow\bar{\xi}+C$, which shifts the range to
$-C\leq\bar{\xi}<\infty$.

\noindent
{}~ {}~ {}~ {}~ \textbf{Case (a).} Let us first consider $\alpha<1$, i.e., the region
I in Fig.\ref{regimes}, where $V_{+}$ is replaced by $\left|V_{+}\right|$.
In this case we have to match the expanding true AdS vacuum solution
with radius $R_{-}$: 
\begin{equation}
\varrho\left(\bar{\xi}\right)=\frac{2R_{-}\left(\bar{\xi}+C\right)}{\left|1-\left(\bar{\xi}+C\right)^{2}\right|}\ ,\label{eq:62}
\end{equation}
with the expanding false vacuum solution with $R_{+}>R_{-}$, given
by (\ref{eq:60}). The matching point $\bar{\xi}_{b}$, and the constant
$C$ are determined by the fact that $\varrho$ must be continuous
at $\bar{\xi}_{b}$, and its derivatives must satisfy (\ref{eq:26aa}),
from which it follows that $\dot{\varrho}_{+}<\dot{\varrho}_{-}$.

\begin{figure}[h!]
\centering 
\includegraphics[width=0.5\textwidth]{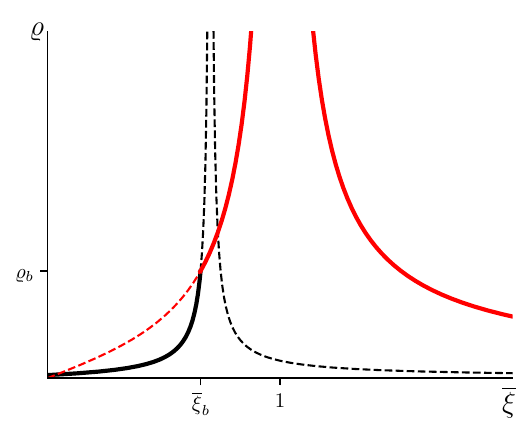} \caption{$\varrho$ versus $\bar{\xi}$ when $\alpha<1$:
matching the expanding true AdS vacuum solution with the expanding false vacuum solution at the matching point $\bar{\xi}_{b}$.}
\label{Fig3afinal} 
\end{figure}

As can be seen from Fig.\ref{Fig3afinal}, in this case $\bar{\xi}_{b}<1-C$
and $C>0$. It follows from (\ref{eq:18aa}), using (\ref{eq:8aa}),
that the contribution of the false AdS vacuum in the final configuration,
which is calculated by subtracting the action for the bubble with
radius $\varrho_{b}$ with potential $V_{+}$ from the initial false
vacuum action, is: 
\begin{equation}
S_{+}^{f}=S^{i}+2\pi^{2}V_{+}\intop_{0}^{\bar{\xi}_{b}}\frac{\varrho^{4}}{\bar{\xi}}d\bar{\xi}+\frac{6\pi^{2}}{\kappa}\varrho_{b}^{2}\dot{\varrho}_{+}=S^{i}+\frac{12\pi^{2}}{\kappa^{2}V_{+}}\left(1-\dot{\varrho}_{+}^{3}\right)\ ,\label{eq:63}
\end{equation}
where 
\begin{equation}
\dot{\varrho}_{+}=+\sqrt{1+\frac{\varrho_{b}^{2}}{R_{+}^{2}}}\ .\label{eq:64}
\end{equation}

Similarly, we obtain the contribution of the true AdS vacuum inside
the bubble to the final action: 
\begin{equation}
S_{-}^{f}=-2\pi^{2}V_{-}\intop_{-C}^{\bar{\xi}_{b}}\frac{\varrho^{4}}{\left(\bar{\xi}+C\right)}d\bar{\xi}-\frac{6\pi^{2}}{\kappa}\varrho_{b}^{2}\dot{\varrho}_{-}=-\frac{12\pi^{2}}{\kappa^{2}V_{-}}\left(1-\dot{\varrho}_{-}^{3}\right)\ ,\label{eq:65}
\end{equation}
where 
\begin{equation}
\dot{\varrho}_{-}=+\sqrt{1+\frac{\varrho_{b}^{2}}{R_{-}^{2}}}\ .\label{eq:67}
\end{equation}
Finally, if these results are combined with the wall contribution
to the final action $S_{w}^{f}=2\pi^{2}\sigma\varrho_{b}^{3}$, the
same expression (\ref{eq:40ab}) is obtained: 
\begin{equation}
S^{f}-S^{i}=\frac{2\pi^{2}\sigma\varrho_{b}^{3}}{\left(1+\dot{\varrho}_{+}\right)\left(1+\dot{\varrho}_{-}\right)}\ ,\label{eq:68}
\end{equation}
where $\sigma$ is the surface tension of the bubble and $\varrho_{b}$,
$\dot{\varrho}_{+}$ and $\dot{\varrho}_{-}$ are given in (\ref{eq:57}),
(\ref{eq:64}) and (\ref{eq:67}), respectively.

According to (\ref{eq:57}), the radius of the critical bubble in
\textit{region} I in Fig.\ref{regimes} can change from $3\sigma/\varepsilon$
to infinity, as we approach the edge of this region. We obtain the
results of standard field theory (\ref{eq:42abc}) and (\ref{eq:45ac}),
when $\varrho_{b}\ll R_{+},R_{-}$. For the bubble with the radius
$R_{+}\gg\varrho_{b}\gg R_{-}$, gravity plays a significant role
and the action (\ref{eq:68}) becomes: 
\begin{equation}
S^{f}-S^{i}\simeq\pi^{2}\sigma\varrho_{b}^{2}R_{-}\ .\label{eq:69}
\end{equation}
If $\varrho_{b}\gg R_{+},R_{-}$ then, 
\begin{equation}
S^{f}-S^{i}\simeq2\pi^{2}\sigma\varrho_{b}R_{+}R_{-}\ ,\label{eq:70}
\end{equation}
and it becomes infinite when 
\begin{equation}
\varepsilon\rightarrow\frac{3}{4}\kappa\varphi_{0}^{2}V_{bar}\left(1+\sqrt{\frac{\left|V_{+}^{*}\right|}{V_{bar}}}\right)\ .\label{eq:71}
\end{equation}

\noindent
{}~ {}~ {}~ {}~ \textbf{Case (b).} Let us now turn to the case $\alpha>1$, i.e., the
region III in Fig.\ref{regimes}, where $V_{+}$ is replaced by $\left|V_{+}\right|$,
corresponding to 
\[
\frac{3}{4}\kappa\varphi_{0}^{2}V_{bar}\left(1-\sqrt{\frac{\left|V_{+}^{*}\right|}{V_{bar}}}\right)>\varepsilon>0.
\]

\begin{figure}[h!]
\centering 
\includegraphics[width=0.5\textwidth]{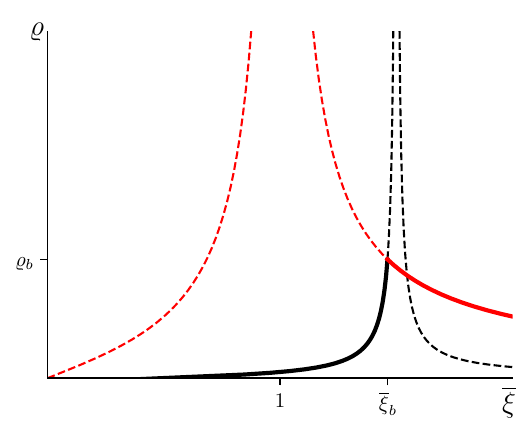} \caption{$\rho$ versus $\bar{\xi}$ when $\alpha>1$:
matching the expanding true AdS vacuum solution with the contracting false vacuum solution at the matching point
$\bar{\xi}_{b}$.}
\label{Fig3bfinal} 
\end{figure}

In this case, the solution for the critical bubble exists and we have
to calculate the action with this solution. As can be seen from  
Fig.\ref{Fig3bfinal}, we need to match the expanding branch of the true
AdS vacuum with the contracting branch of the false AdS vacuum at
the point $1+\left|C\right|>\bar{\xi}_{b}>1$, where $C$ in (\ref{eq:62})
must be negative. To calculate the contribution of the false AdS vacuum
to the action after the bubble formation, we proceed as before, and
subtract from the initial action the action for the false vacuum bubble
with the radius $\varrho_{b}$. In this case, there is a singularity
at $\bar{\xi}=1$, and to regularize this singularity, we divide the
integration domain into two subdomains $(1-\delta)^{1/2}\geq\bar{\xi}\geq0$,
and $\bar{\xi}_{b}\geq\bar{\xi}\geq(1-\delta)^{1/2}$, where $\delta\ll1$.
The result is: 
\begin{align}
 & S_{+}^{f}=S^{i}+2\pi^{2}V_{+}\left[\intop_{0}^{(1-\delta)^{1/2}}\frac{\varrho^{4}}{\bar{\xi}}d\bar{\xi}+\intop_{(1+\delta)^{1/2}}^{\bar{\xi}}\frac{\varrho^{4}}{\bar{\xi}}d\bar{\xi}\right]+\frac{6\pi^{2}}{\kappa}\left[\left.\varrho^{2}\dot{\varrho}\right|_{0}^{(1-\delta)^{1/2}}+\left.\varrho^{2}\dot{\varrho}\right|_{(1+\delta)^{1/2}}^{\bar{\xi}_{b}}\right]\nonumber \\
 & =S^{i}-\frac{24\pi^{2}}{\kappa^{2}V_{+}}\left(\frac{8+6\delta^{2}}{\delta^{3}}\right)+\frac{12\pi^{2}}{\kappa^{2}V_{+}}\left(1+\dot{\varrho}_{+}^{3}\right)\ ,\label{eq:72}
\end{align}
where $\dot{\varrho}_{+}$ is defined in (\ref{eq:64}). The contributions
of the true vacuum and the wall are both finite. If we consider that
$V_{+}$ is negative, we see that $S^{f}-S^{i}$ tends to $+\infty$
when $\delta\rightarrow0$, and thus the probability of transitions
for $\alpha>1$ (region III in Fig.\ref{regimes}) vanishes, although the corresponding
solution for the bubble exists. From this we conclude, that for a
given $V_{+}$ and $V_{bar}$, the AdS false vacuum is stable if 
\begin{equation}
\varepsilon=V_{+}-V_{-}<\frac{3}{4}\kappa\varphi_{0}^{2}V_{bar}\left(1+\sqrt{\frac{\left|V_{+}^{*}\right|}{V_{bar}}}\right)\ ,\label{eq:73}
\end{equation}
despite the presence of a deeper true vacuum $V_{-}<V_{+}.$

\section{Discussion}

Working within the thin-wall approximation, we have shown that the
false dS vacuum is always unstable, irrespective of the height of
the barrier $V_{bar}$, which separate the false dS vacuum with positive
potential $V_{+}$ from the true vacuum with potential $V_{-}$. The
true vacuum can be either a dS vacuum with a smaller potential, a
Minkowski vacuum or an AdS vacuum. This conclusion is not very intuitive
from field theory perspective. Namely, if the radius of the bubble
obtained by neglecting gravity, $\varrho_{b}=3\sigma/\varepsilon$,
starts to exceed the dS radius of the false vacuum $R_{+}$, one might
naively expect that tunneling out of the false vacuum cannot take
place, and it becomes stable despite the presence of the true vacuum
with a smaller potential. This can happen, for example, if, either
the surface tension $\sigma\sim\varphi_{0}V_{bar}^{1/2}$ is too large
or the distance between the depths of the vacua $\varepsilon=V_{+}-V_{-}$
is too small.

However, as we have seen, gravity becomes very important in such circumstances
and the negative contribution of gravity to energy cannot be neglected,
when either the vacua are almost degenerate, i.e., $\varepsilon\ll\varphi_{0}\sqrt{\kappa V_{+}V_{bar}}$
for $V_{bar}<V_{+}/\left(\kappa\varphi_{0}^{2}\right)$, or if the
height of the barrier exceeds $V_{+}/\left(\kappa\varphi_{0}^{2}\right)$
independently of $\varepsilon$. In the first case, the vacuum decay
occurs via critical bubbles with a radius of the order of the false
dS radius $R_{+},$ and the decay rate is determined by the action
(\ref{eq:47ac}), while in the second case for a very high potential
barrier the size of the bubble can be much smaller than the dS radius
and the decay rate is given by the action (\ref{eq:50ac}), which
depends only on the potential in the false vacuum. Since $V_{bar}$
cannot exceed the Planck scale, the minimal size of these bubbles
is about $l_{Pl}/\sqrt{\kappa\varphi_{0}^{2}}\sim 1/\varphi_{0}$. It should be noted
that the decay rate via gravitationally dominated instantons is exponentially
small for the small initial cosmological constant, and vanishes when
$V_{+}$ approaches zero.

In contrast, the Minkowski or AdS false vacua are only unstable if
the relative depth of the true vacuum exceeds the critical values given in  (\ref{eq:56})
or (\ref{eq:73}). Otherwise, the false vacuum is stable despite the
existence of the true vacuum with $V_{-}<V_{+}$.

\section*{Acknowledgements}

The work of V. M. is supported in part by the Deutsche Forschungsgemeinschaft
(DFG, German Research Foundation) under Germany's Excellence Strategy
-- EXC-2111 -- 390814868. The work of Y.~O. \ is supported in
part by the Israel Science Foundation Excellence Center, the US-Israel
Binational Science Foundation, and the Israel Ministry of Science.
The work of A. S. is supported in part by the Center for Integration in Science of the 
Israel Ministry of Aliyah and Integration.

\bibliographystyle{JHEP}
\bibliography{literature}

\end{document}